# Pressure-Induced Insulator-Metal Transition in Silicon Telluride from First-Principles Calculations

*Romakanta Bhattarai and Xiao Shen*

Department of Physics and Materials Science, University of Memphis, Memphis, TN, 38152

**ABSTRACT**

Silicon telluride ($Si_2Te_3$) is a two-dimensional semiconductor with unique structural properties due to the size contrast between Si and Te atoms. A recent experiment shows that the material turns metallic under hydrostatic pressure, while the lattice structure of the metallic phase remains to be identified. In this paper, we propose two metallic phases, M1 and M2, of $Si_2Te_3$ using the evolution algorithm and first-principles density functional theory (DFT) calculations. Unlike the presence of Si-Si dimers in the semiconducting (SC) phase, both M1 and M2 phases have individual Si atoms, which play important roles in the metallicity. Analysis of structural properties, electronic properties, dynamical as well as thermal stability is performed. The energies of these new structures are compared with the SC phase under the subsequent hydrostatic pressure up to 12 GPa. The results show that M1 and M2 phases have lower energies under high pressure, thus elucidating the appearance of the metallic phase of $Si_2Te_3$. In addition, the external pressure causes the SC phase to have an indirect-direct-indirect bandgap transition. Analysis of Raman spectra of the SC phase at a different pressure shows the shifting of the major Raman peaks, and finally disappearing confirms the phase transition. The results are in good agreement with the experimental observations. The understanding of the insulator-metal phase transition increases the potential usefulness of the material system.

## I. INTRODUCTION

Silicon telluride ($Si_2Te_3$) is a two-dimensional material with a complex structural configuration. This material has been successfully fabricated into a few atomic layers thickness in the past several years[1–4]. As Si atoms have relatively small sizes, they form dimers and fill only 2/3 of the allowed 'metal' sites between the hexagonal close-packed Te layers, thereby leaving the 1/3 sites vacant. The dimer orientation is possible along both in-plane and out-of-plane directions, and rotation of the dimer has an activation energy of 1 eV, indicating that it can happen at room temperature[5]. The presence of Si-Si dimers and their possible orientations adds extra interest to the material as the dimer orientations alter the fundamental properties of the material and impact potential applications.

$Si_2Te_3$ is one of the rare natural p-type 2D semiconductors[3] and also possesses a set of interesting properties. Theoretical investigation shows that $Si_2Te_3$ is one of the most flexible 2D materials[6]. First-principles calculations and experiments show that it possesses anisotropic optical properties [7,8]. Experiments by Wang et al. show its optoelectronic properties can be modified through doping and intercalation of Ge and Cu without altering the fundamental host lattice[9]. Other members of the Si-Te family, such as SiTe and $SiTe_2$, are also confirmed and have shown interesting properties[10–19]. A recent theoretical investigation discovered a novel structure of $SiTe_2$[20] that is more stable than the $CdI_2$ type that is common in $MX_2$ compounds, which again manifests the uniqueness of this material system.

A recent experiment by Johnson et. al[21] shows that $Si_2Te_3$ nanoplates undergo a color change from red to black around a hydrostatic pressure of 9.5 GPa, indicating a phase transition to a metallic phase. Such phase transition can be observed at a relatively lower pressure around 7 GPa if Mn is intercalated between the layers. Also, the major Raman active mode in the SC phase at 144 $cm^{-1}$ shifts under external pressure and finally disappears, which could be the sign of phase transition. The crystal structure of the metallic $Si_2Te_3$ remains to be explored. In another recent experiment, Wu et al. discovered that $Si_2Te_3$ nanowires could be switched between the semiconducting and metallic states under external electric voltage[22]. In addition, phase transition around 673-723 K was reported by Bailey[23], Ziegler[24], and Gregoriades[2], which is associated with silicon dimers break up. These pressure-, electrical-, and temperature-induced phase changes make it particularly interesting to investigate other possible phases of $Si_2Te_3$, especially the metallic ones.

In this paper, we report a computational study of $Si_2Te_3$ under high pressure using the evolutionary algorithm combined with first-principles density functional theory (DFT) calculations. We propose two candidate metallic phases of $Si_2Te_3$ to account for the experimental observation at high-pressure. The total energies of the metallic phases are compared with the original semiconducting phase under different external pressure, from which the transition pressures are obtained and are in good agreement with experiments. We also find that the semiconducting $Si_2Te_3$ undergoes a bandgap transition from indirect to direct and back to indirect with the pressure. Analysis of Raman spectra shows the positive shifting of the major peak at 144 $cm^{-1}$ with the pressure and finally disappearing, which agrees with the experiments.

## II. COMPUTATIONAL DETAILS

We use the universal crystal structure predictor USPEX[25,26] to investigate possible stable phases of $Si_2Te_3$. The predictor works under the evolutionary algorithm coupled with the first-principles density functional theory method. Identification of the global free-energy minimum is required to predict the most stable phase among all possible structures generated. A total of 30 structures are considered as the initial population, which are generated randomly by using the space group symmetry. The population in each generation is kept fixed throughout the search with the evolutionary algorithm. The stopping criterion for the evolutionary algorithm search is that the most-fit structure does not change for 15 generations. We choose 60% of the current generation to produce the next generation. In total, 50% of the population is produced by heredity, 20% produced randomly using the space group symmetry, 10% produced by soft mutation, 10% produced by lattice mutation, and the remaining 10% produced by permutation in each generation afterward. The search is carried out under an external hydrostatic pressure of 10 GPa. During the evolutionary algorithm search, first-principles DFT calculations using VASP (Vienna Ab initio Simulation Package)[27] package with the Perdew-Burke-Ernzerhof (PBE) exchange-correlation funcational[28] and the projected augmented wave method[29] are used to relax the structures and obtain the total energies. The maximum kinetic energy cutoff is 320 eV. The convergence criteria for electronic and ionic relaxation are respectively $10^{-5}$ eV and $10^{-4}$ eV, along with.

For accurate comparisons of the energies, the low energy structures obtained by the evolutionary algorithm are optimized further with the VASP with tighter numerical settings. Plane-wave basis set with a kinetic energy cutoff of 500 eV is used for the expansion of wavefunction. The criteria for electronic and force convergence are set to $10^{-9}$ eV and $10^{-4}$ eV/Å, respectively. Gamma centered k-point grids of 9×9×5, 9×9×9, and 13×13×5 are used in the integration of Brillouin zones for the semiconducting (SC) phase and two metallic phases, respectively. The calculations use both PBE[28] and PBEsol[30] versions of the exchange-correlation functional. The dynamical stability of both metallic structures is confirmed by the phonon spectra obtain using Phonopy package[31], which uses the finite displacement method with supercells. The dynamical stability is further verified by the ab initio molecular dynamics simulations using the supercells containing 160 atoms and 90 atoms for M1 and M2 phases at 500 K for 30 ps with a time step of 2 fs.

The Raman spectra of the SC phase under different pressures are calculated using the density functional perturbation theory (DFPT)[32]. We use the Quantum Espresso package[33] with the norm-conserving pseudopotential generated via Rappe-Rabe-Kaxiras-Joannopoulos method[34]. The exchange-correlation functional of PBE type is used in the calculations. Plane-wave basis sets with the cutoff energy of 80 Rydberg (Ry) is used for the expansion of wave function. The energy and force are converged to $10^{-8}$ Ry and $10^{-7}$ Ry/Bohr, respectively, during the geometrical optimization. Brillouin zone is sampled with a 9×9×5 k-points grid centered at Gamma point. The total energy is converged to $10^{-12}$ Ry and $10^{-14}$ Ry in self-consistent and phonon calculations, respectively. Finally, Raman coefficients are computed from the second-order response function[35].

## III. RESULTS AND DISCUSSION

The global search for the low energy configurations of $Si_2Te_3$ using the evolutionary algorithm with USPEX and the density functional theory is done for the unit cells consisting of 5 to 80 atoms. During the search, an external pressure of 10 GPa is applied to explore the stable high-pressure phases of $Si_2Te_3$. A total of 2398 structures are generated, among which we find two structures with the lowest energy: one with ten atoms per unit cell and another with five atoms per unit cell, which are named as M1 and M2 phases. We compare the energy per atom on both the phases under PBE and PBEsol functionals. Under PBE functional, energy per atom of M1 and M2 are -3.91 eV & -3.90 eV at 10 GPa respectively. However, under PBEsol, the respective values are -3.30 eV & -3.32 eV at 10 GPa. The very small energy difference is within the error bars associated with the exchange-correlation functional. Thus, we can only conclude that they are close in energy and could both exist in nature.

The crystal structures of the M1 and M2 phases are shown in Figure 1 and Figure 2, respectively. The corresponding lattice parameters of M1 and M2 are shown in Table I. The M1 phase has a hexagonal crystal lattice, whereas the M2 has a triclinic crystal lattice. We also explored ABAB and ABCABC types of stacking of the layers in the M1 phase. It is found that the energies of different stacking only differ by less than 1 meV. Therefore, different stackings are likely to occur in bulk and multiplayer of the M1 phase. In this work, we focus on the configuration shown in Fig. 1. To the best of our knowledge, these two structures have not been reported in other 2D chalcogenides.

Table I: Structural parameters of the predicted M1 and M2 phases of $Si_2Te_3$

| Phase | a (Å) | b (Å) | c (Å) | α (°) | β (°) | γ (°) | Bravais lattice |
|---|---|---|---|---|---|---|---|
| M1 | 6.82 | 6.82 | 7.48 | 90 | 90 | 60 | Hexagonal |
| M2 | 3.90 | 3.90 | 10.93 | 100.26 | 90.07 | 120 | Triclinic |

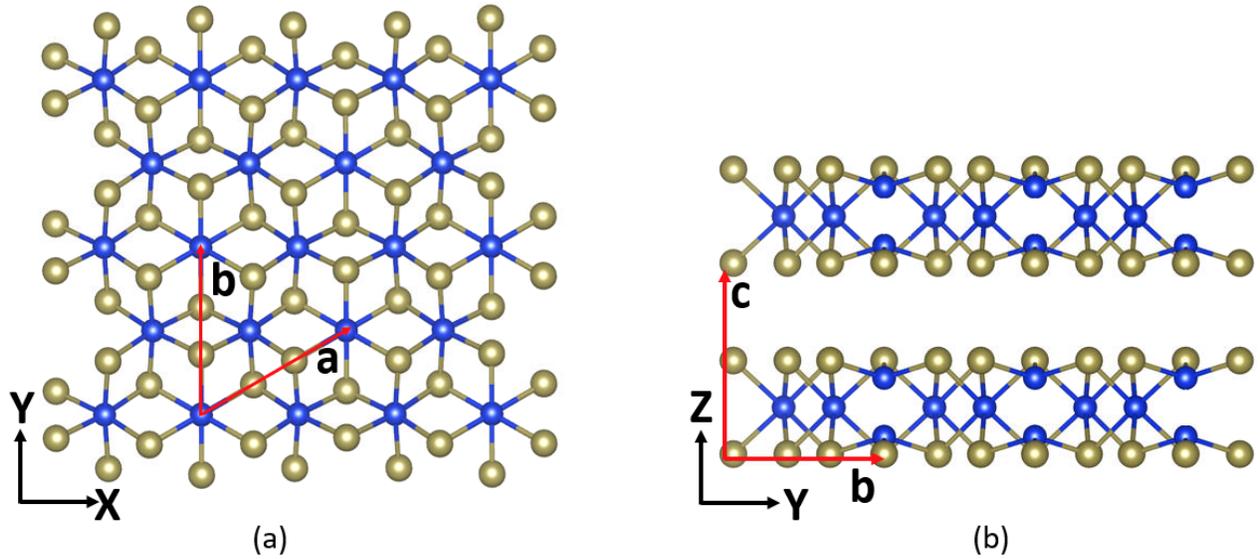

Figure 1: (a) Top and (b) side views of M1 phase of $Si_2Te_3$ with lattice vectors marked by solid red lines. Te and Si atoms are represented in tan and blue color, respectively.

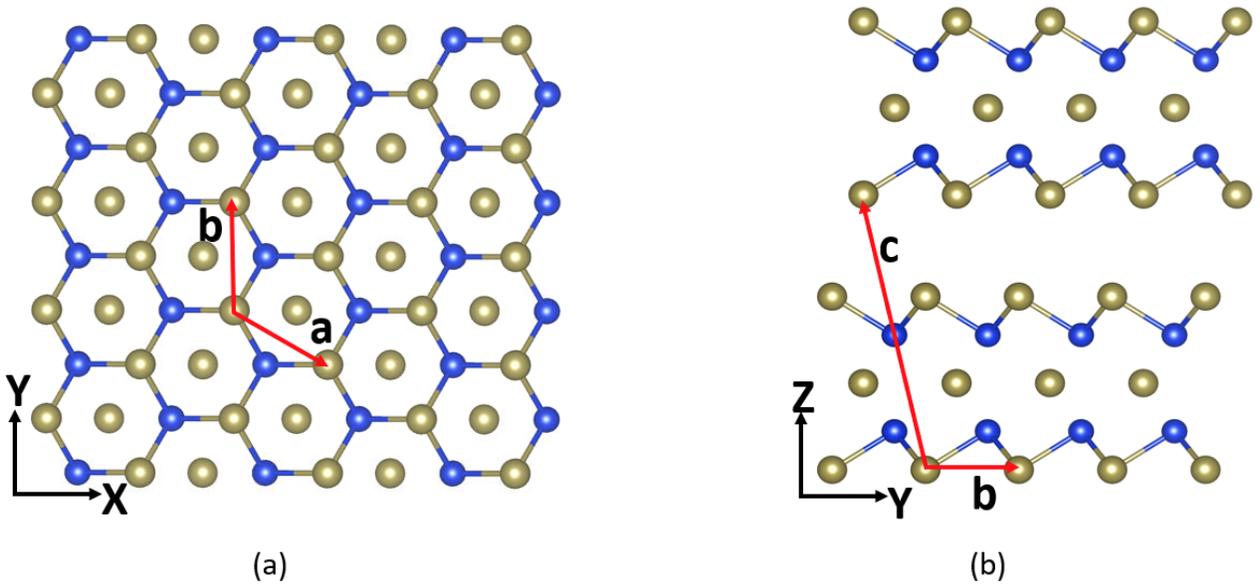

Figure 2: (a) Top and (b) side views of the M2 phase of $Si_2Te_3$.

The dynamical stabilities of M1 and M2 are tested under phonon dispersion calculation using the finite displacement method with supercells of 160 atoms at 10 GPa pressure. The phonon spectra have no imaginary phonon modes, as shown in Figure 3(a) and 4(a), confirming their dynamical stabilities. We further confirm their dynamical stability by performing the ab-initio molecular dynamic (AIMD) simulation of both phases at 500 K and 10 GPa, which are shown in Figures 3(b) and 4(b) for the M1 and M2 phases, respectively. No structure reconstruction is found over the entire 30 ps. The fluctuation of the total energy per atom during the simulation time of 30 ps without any sudden drop confirms that both the phases are dynamically stable at 10 GPa pressure.

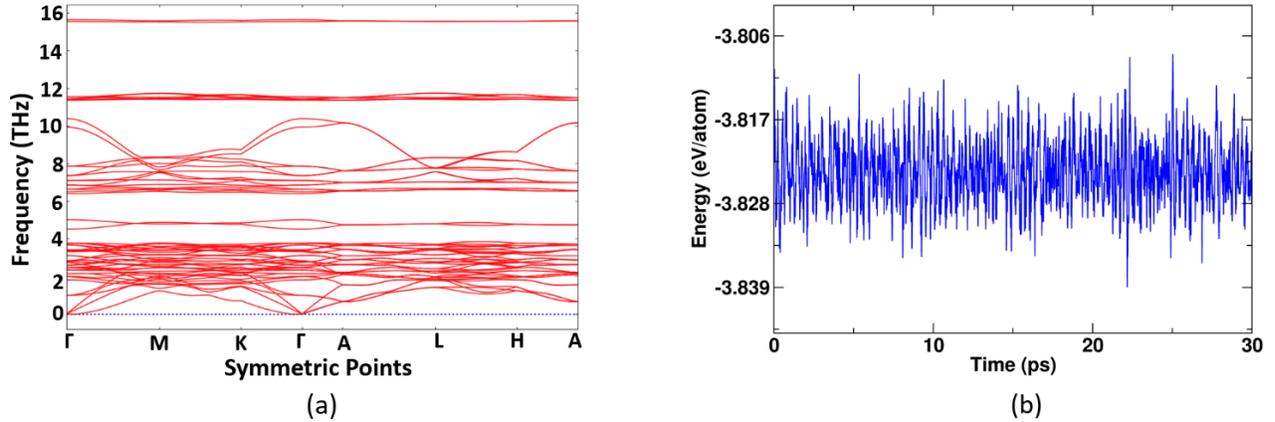

Figure 3. (a) Phonon band structures of the M1 phase. (b) Energy of the M1 phase during AIMD simulation at 500 K over 30 ps.

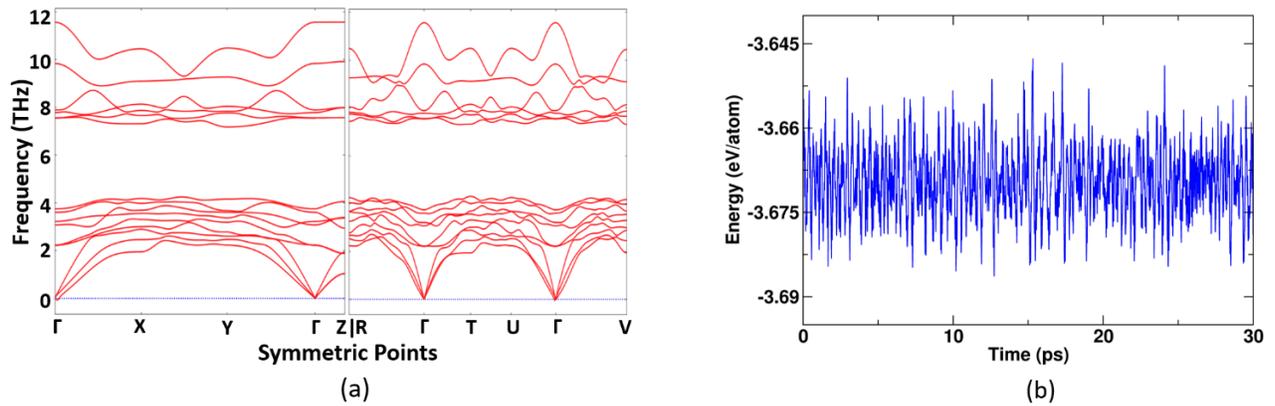

Figure 4. (a) Phonon band structures of the M2 phase. (b) Energy of the M2 phase during AIMD simulation at 500 K over 30 ps.

Next, we discuss the electronic properties of the M1 and M2 phases. Figure 5 shows the electronic band structures and densities of states (DOS) obtained using DFT with the spin-orbit interaction. From the band structure of the M1 phase (Figure 5(a)), we can see there is no bandgap, indicating the phase is metallic. The corresponding partial and total DOS of M1 is plotted in Figure 5(b). It shows that the p-orbital of the

Te atom mainly constitutes the valence band, whereas the p-orbital of both the Si and Te atom have major contributions to the conduction band. There is a small contribution also coming from the s-orbital of Si and Te atoms as well as the d-orbital of Te atom. Similar patterns are seen in Figures 5(c) and 5(d) for the M2 phase, where both the band structures and DOS indicate the metallic nature. As the DFT method is known to underestimate the band gap, to further verify the metallic nature of the M1 and M2 phases, we perform hybrid DFT (HSE06[36,37]) calculation. Both the band structures and DOS from the HSE method (see Supplemental Information) confirm that M1 and M2 are metallic.

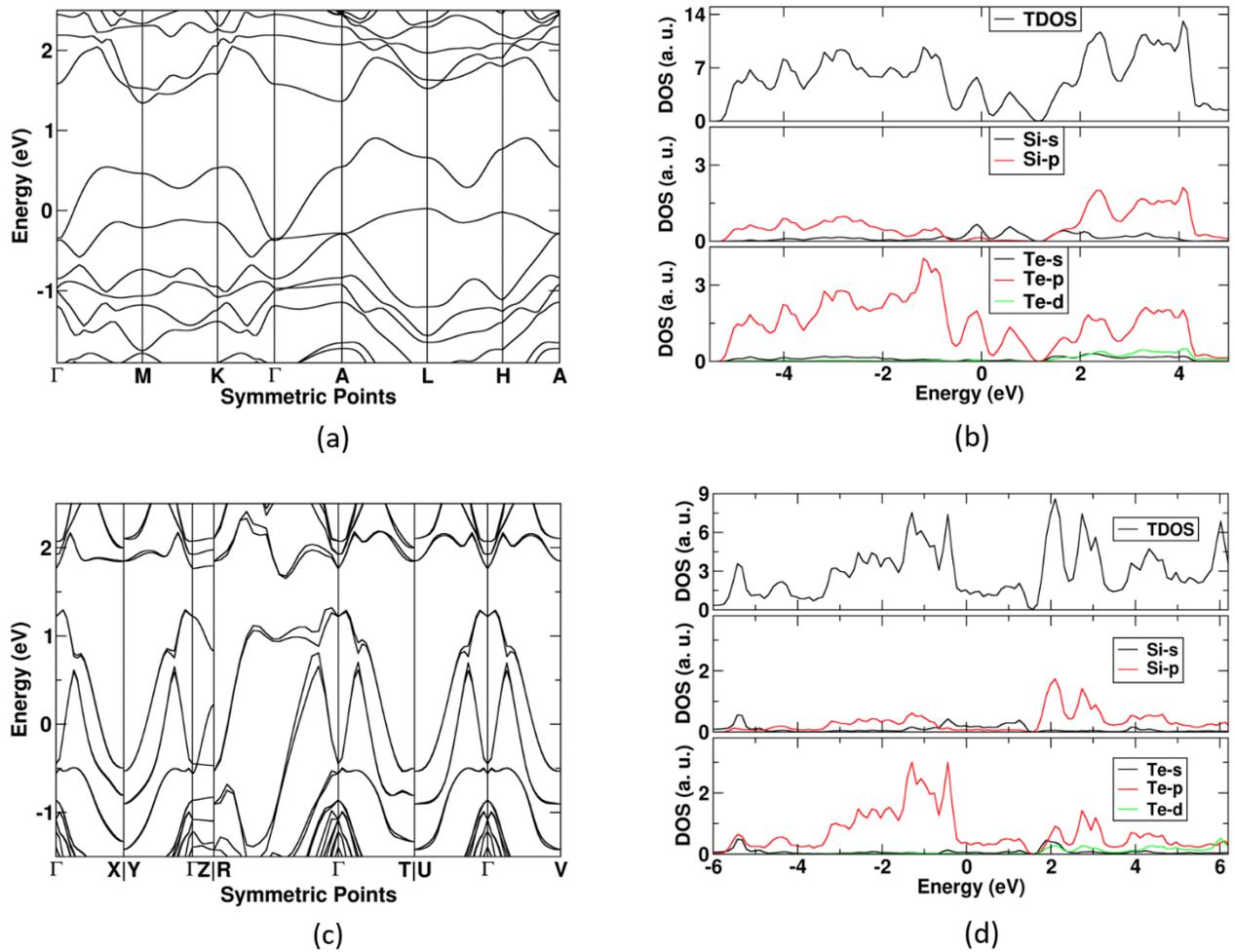

Figure 5: (a) Electronic band structures of the M1 phase along the high symmetric points in Brillouin zone (b) Partial and total DOS of the M1 phase. (c) Electronic band structures of the M2 phase along the high symmetric points in Brillouin zone. (d) Partial and total DOS of the M2 phase. Effect of spin orbit coupling is considered in the calculations.

To get a deeper insight into the electronic properties of $Si_2Te_3$, we perform the Bader charge analysis on the M1 and M2 phases and compare them with the result from the well-known SC phase. In the M1 phase, we found that the amount of charge transferred from each Si atom is not the same. A charge of 0.523e is transferred from half of the Si atoms (labeled as Si1 in Figure 6(a)), whereas 0.459e is transferred from other halves (labeled as Si2 in Figure 6(a)). However, the amount of charge received by each Te atom is the same, which is 0.327e. This kind of transfer mechanism is found to be opposite in the M2 phase, where a total of 0.483e charge gets transferred from each Si atom, but the amount of charge gained by all Te atoms is not the same throughout the cell. Each Te atom at the surface gets 0.339e charge, whereas a Te atom in the middle layer gets relatively less charge (0.288e) as compared to the surface atom. The higher charge in Te atoms at the surface is higher than in the middle layer, making the surface more conducting. In the semiconducting phase, each Si atom transfers a 0.609e charge, whereas each Te atom gets a 0.406e amount of charge.

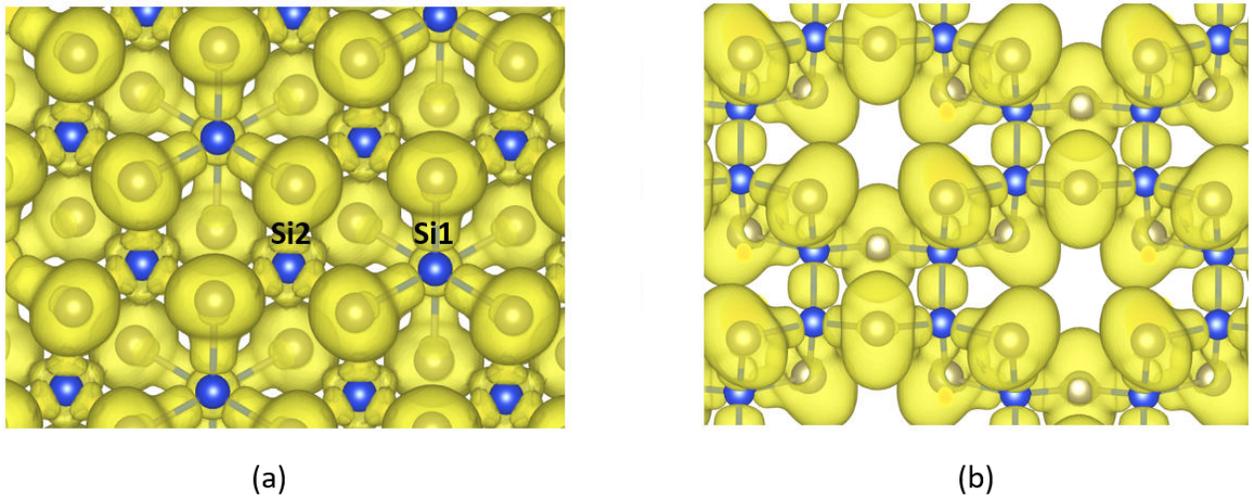

(a)        (b)

Figure 6. Top views of the electron localization function (ELF) in the M1 (a) and SC (b) phases of $Si_2Te_3$. The iso-surface level is 0.72 in each case.

To explain the charge transfer mechanism in various phases, we analyze the chemical bonding in $Si_2Te_3$ using the electron localization function (ELF). Figures 6(a) and 6(b) are top views of the M1 and SC phases, respectively. In Fig. 6(a), we observe two different types of bonding in the M1 phase: every Te atom has a lone pair and participates in one Si-Te bond that indicates covalent bonding. However, only half of the Si atoms (marked as Si1 in Fig. 6(a)) participates in the Si-Te bonds. The remaining half of the Si atoms (marked as Si2) are not bonded to Te. This is consistent with the different amounts of charge transferred pates from two types of Si atoms, as discussed above. Meanwhile, in the SC phase, each Si atom participates in one Si-Si bond and three Si-Te bonds, and each Te atom participates in two Si-Te bonds and possesses one lone pair, as shown in Figure 6(b). This result matches with the discussion above that all Si atoms in the SC phase are losing charge equally, and all Te atoms are gaining charge equally.

Next, we analyze the ELF in the M2 phase and compare it with that in the SiTe structure. The reason for such comparison is that the top and bottom surfaces of the M2 phase of $Si_2Te_3$ is similar to SiTe. Figure 7(a) and 7(b) are the side views of the M2 phase and SiTe, respectively. In Figure 7(a), it is clearly seen the upper and lower Te atoms (i.e., surface atoms) feature lone pair and Si-Te bond. Meanwhile, there is no indication of Si-Te bonds in the middle Te layer. The ELF profiles of the top and bottom layers are very similar to that of the bulk SiTe structure, which is shown in Figure 7(b). This similarity led us to conclude that the Si atoms are covalently bonded with the surface Te atoms. Meanwhile, as the Te atoms in the middle are slightly negatively charged, they likely form ionic bonding with the positive Si atoms. Such a layer of Te atoms may have interesting electronic transport properties (see Supplemental Information for the charge density at the Fermi level).

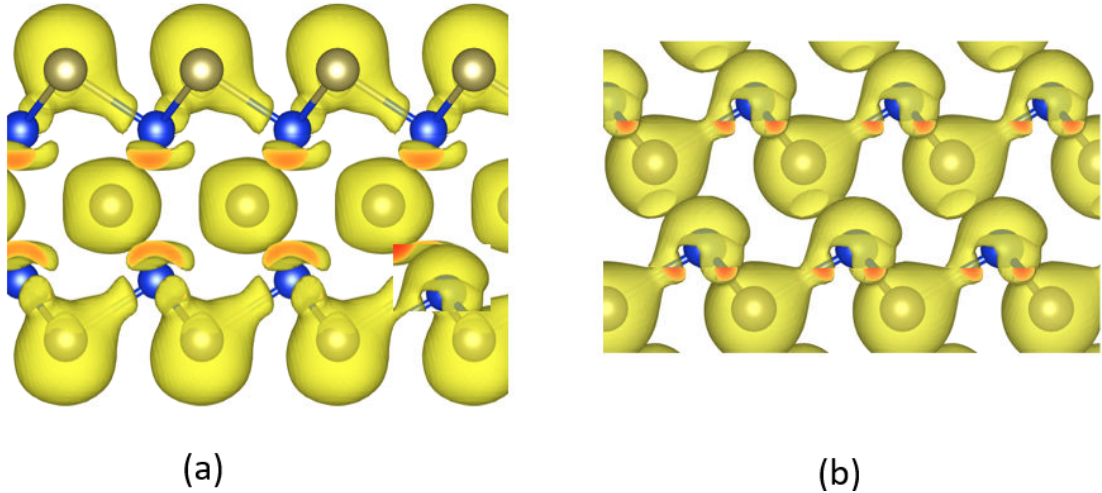

Figure 7. Side views of ELF in the M2 phase of $Si_2Te_3$ (a) and SiTe (b) The iso-surface level is 0.73 in each case.

## IV. Effect of External Pressure

We now discuss the effect of external hydrostatic pressure on the SC phase of $Si_2Te_3$. First, we calculate the phase transition pressure. We take all three phases viz; the M1, M2, and SC phases of $Si_2Te_3$ and apply a subsequent hydrostatic pressure up to 12 GPa. The energies per atom of each phase in the pressure range of 0-12 GPa are shown in Figure 8(a). Initially, when the pressure is not applied, the total energy per atom of M1, M2, and SC phases are -4.04, -4.01, and -4.12 eV, respectively. As the SC phase is the lowest in total energy, it is the ground state at low pressures. However, upon applying the pressure, the energy of the SC phase is found to increases faster as compared to M1 and M2 and exceeds the M1 phase beyond 7.4 GPa and M2 phase beyond 8.7 GPa. It implies that at 7.4 GPa, there is a phase transition from semiconductor to metal and the resulting metallic phase (M1) has the lower energy. Also, at 8.7 GPa, there could be another phase transition from the SC phase to the metallic M2 phase. The transition pressures of 7.4 GPa and 8.7 GPa are in agreement with the experimental value of 9.5 GPa[21]. It could also be possible that the phase observed experimentally was actually either the M1 or the M2 phase or a mixture of the two. Future experiments may resolve this. This possibility of two stable phases of $Si_2Te_3$ adds extra interest to the Si-Te compounds.

We also find that the SC phase undergoes an unusual indirect-direct-indirect band gap transition with the increase of hydrostatic pressure. Figure 8(b) shows the band gap of the SC $Si_2Te_3$ as a function of applied pressure from DFT calculations using the PBE functional, including the spin-orbital interaction. It can be seen that the band gap decreases from 1.48 eV to 0.31 eV as the pressure changes from 0 to 10 GPa. We observe two kinks in the curve that indicate changes in the nature of the band gap. The gap at 0 GPa is 1.48 eV and is indirect. However, the application of hydrostatic pressure changes the band gap to direct at 1.2 GPa. The band gap remains direct until 6.8 GPa, when it changes back to indirect and remains so until the material transitions to the metallic phase. It is interesting to note that a similar pattern of band gap variation and direct-indirect transition with uniaxial strain was also observed in the monolayer $Si_2Te_3$[6].

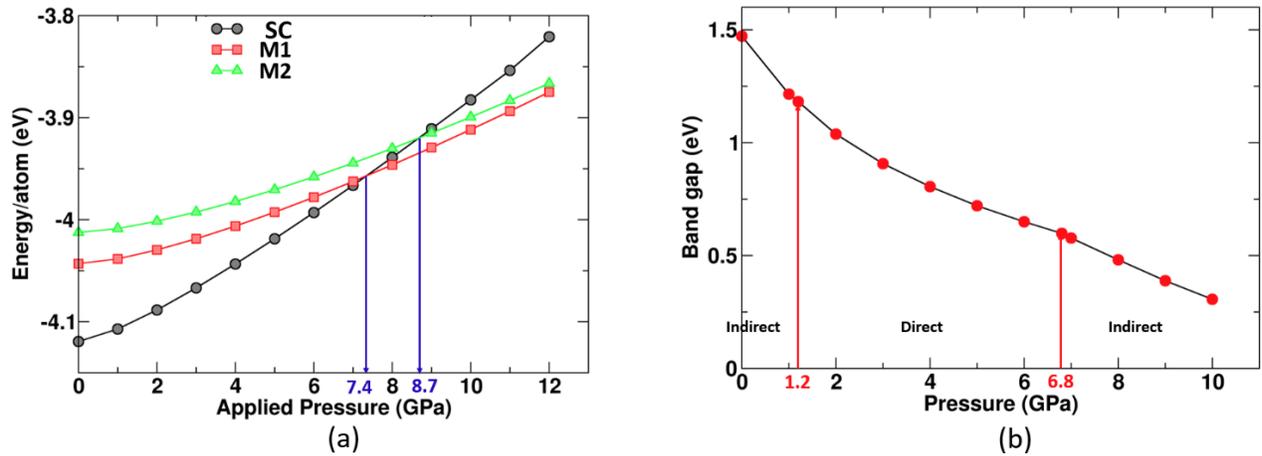

Figure 8: (a) Variation of energy per atom of the SC, M1 and M2 phases of $Si_2Te_3$ (b) DFT-PBE band gap as a function of the applied hydrostatic pressure in the SC phase showing indirect-direct-indirect band gap transitions.

A detailed analysis of the band gap transition is shown in the band structures in Figure 9. At 0 GPa, the VBM of the SC $Si_2Te_3$ lies at the X point in the Brillouin zone, whereas the CBM lies at the Gamma point, resulting in an indirect band gap (Fig 9(a)). As the pressure increases, the CBM shifts towards the X points and arrives at it at 1.2 GPa, thus changing the band gap to direct. The positions of CBM and VBM remain at X points (Fig. 9(b)) up to 6.7 GPa, but as the pressure increase further, the VBM starts to shift towards the Gamma points, causing the band gap to be indirect again (Fig. 9(c)). These shits in CBM and VBM locations indicate that both the electrons and holes in the SC $Si_2Te_3$ are strongly influenced by the pressure.

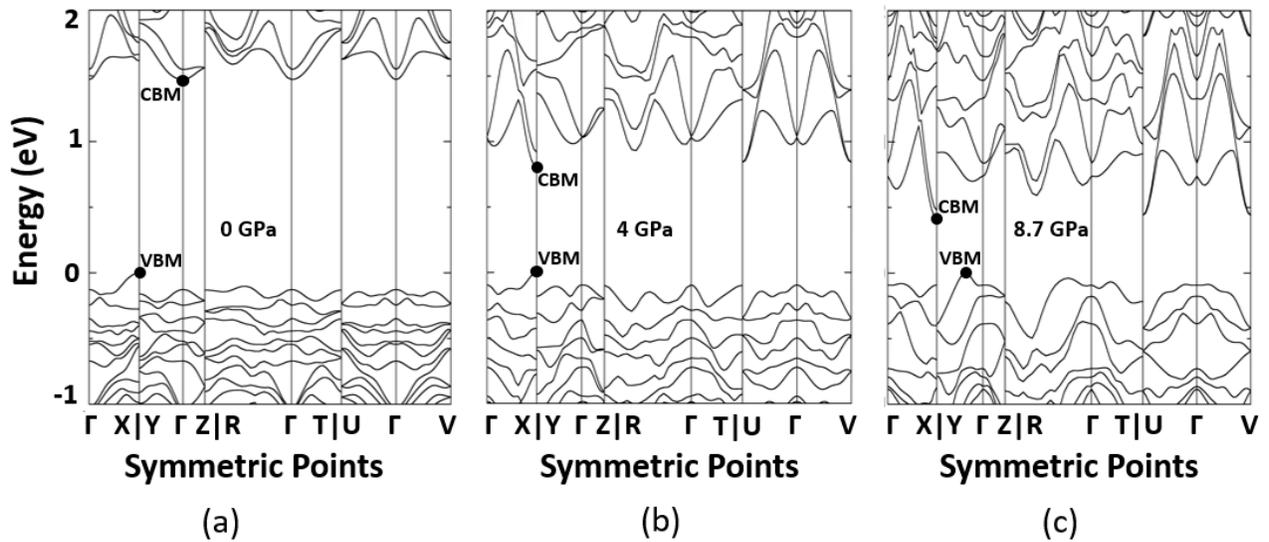

Figure 9: Electronic band structures of the SC phase of $Si_2Te_3$ under DFT with the PBE functional at different pressures showing the gap transition. VBM and CBM refer to the valence band maximum and conduction band minimum. Spin-orbit interaction is considered in the calculation.

In the SC phase of $Si_2Te_3$ under ambient pressure, there is a very strong Raman peak at 144 cm$^{-1}$, which is a well-known $A_{1g}$ mode, along with several other active peaks. Johnson et al. observed that the major Raman peak shifts to higher frequency under external pressure and disappearance at the pressure greater than 10 GPa[21]. To compare with the experiment, we calculate the Raman spectra of the SC phase of $Si_2Te_3$ under different pressures up to 12 GPa. The results are shown in Figure 10. We find a major Raman peak at 144 cm$^{-1}$ at 0 GPa, which agrees well with the experimental value of 144 cm$^{-1}$. The calculated Raman peak shifts towards higher frequencies (wavenumbers) as the pressure increases. At 10 GPa, the peak position has changed to 162 cm$^{-1}$, which is close to the experimental value of 167 cm$^{-1}$ at 9.86 GPa. Overall, the calculations of Raman peak position and the pressure-induced shift are in excellent agreement with experiments.

Next, we discuss the disappearance of Raman peak at pressures greater than 10 GPa. Ideal metals do not have Raman active phonon modes due to the screening from free charge carriers. However, the screening can be incomplete in 2D metals such as the metallic MoS$_2$, where Raman signals have been experimentally observed[38]. The absence of the Raman peak at high pressure in the experiment can be due to either sufficient screening in the metallic nanoplates or the changes in the phonon modes upon the phase transition. Since the M1 and M2 phases are metallic, it is not possible to directly carry out calculations of Raman intensities using the DFPT method. Alternatively, we examine the vibrational modes of the M1 and M2 phases obtained from phonon calculations. No phonon mode is found near the position of the major Raman peak in the SC phase (144 cm$^{-1}$ at 0 GPa and 162 cm$^{-1}$ at 10 GPa). The results suggest that the absence of the A$_{1g}$ mode near 144 cm$^{-1}$ can be used to identify the phase transition to the M1 and M2 phases, even in experiments with smaller and thinner samples.

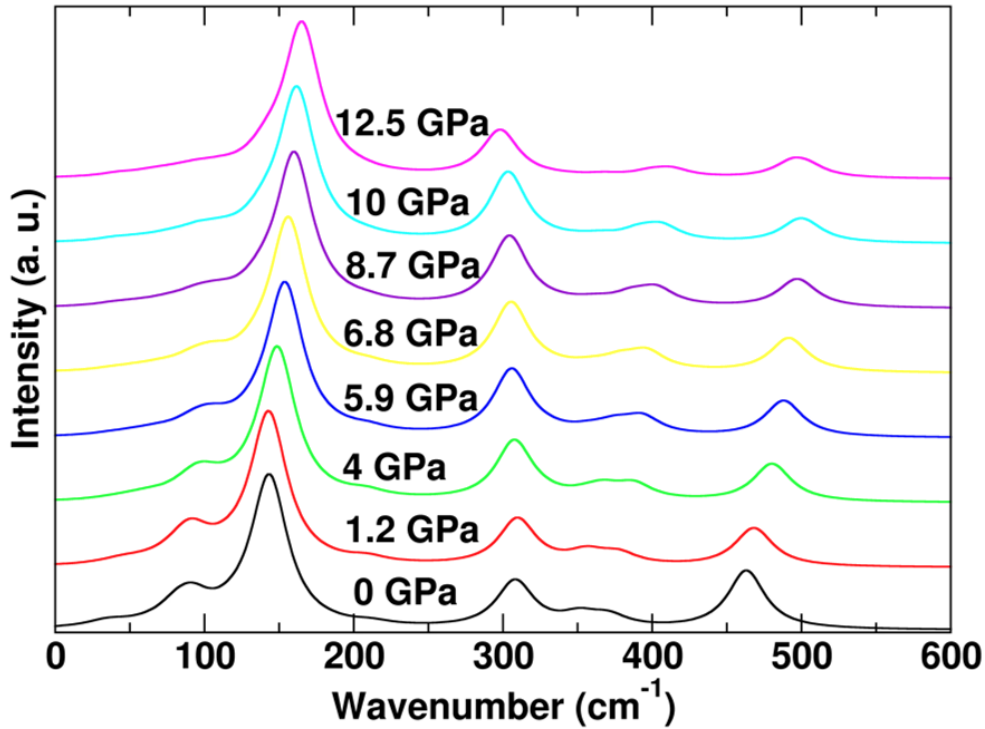

Figure 10: Calculated Raman spectra of the SC phase of Si$_2$Te$_3$ under different pressure.

## V. Conclusion

Two candidate structures, M1 and M2, of the high-pressure metallic phase of $Si_2Te_3$, are discovered using the evolution algorithm and first-principles density functional theory calculations. Analysis of structural, electronic, and vibrational properties is performed. A comparison of energies of new structures with the well-known SC phase as a function of hydrostatic pressure reveals insulator-metal phase transition pressure around 7.4 to 8.7 GPa. Also, the external pressure causes the SC phase to have an indirect-direct-indirect bandgap transition. The calculated Raman spectra of the SC phase shows the shifting of the major peaks towards higher wavenumbers with increasing pressure and finally disappearing upon the phase transition. These results are in agreement with the experimental observations. Overall, the findings highlight the rich structures and properties of $Si_2Te_3$ and the Si-Te system.

**Acknowledgment:**

This work was supported by the National Science Foundation grant # DMR 1709528 and by the Ralph E. Powe Jr. Faculty Enhancement Awards from Oak Ridge Associated Universities (ORAU). The computational resources were provided by the NSF XSEDE grant number TG-DMR 170064 and 170076 and the University of Memphis High-Performance Computing Center (HPCC).

# Supplemental Information

**Pressure-Induced Insulator-Metal Transition in Silicon Telluride from First-Principles Calculations**

*Romakanta Bhattarai and Xiao Shen*

Department of Physics and Materials Science, University of Memphis, Memphis, TN, 38152

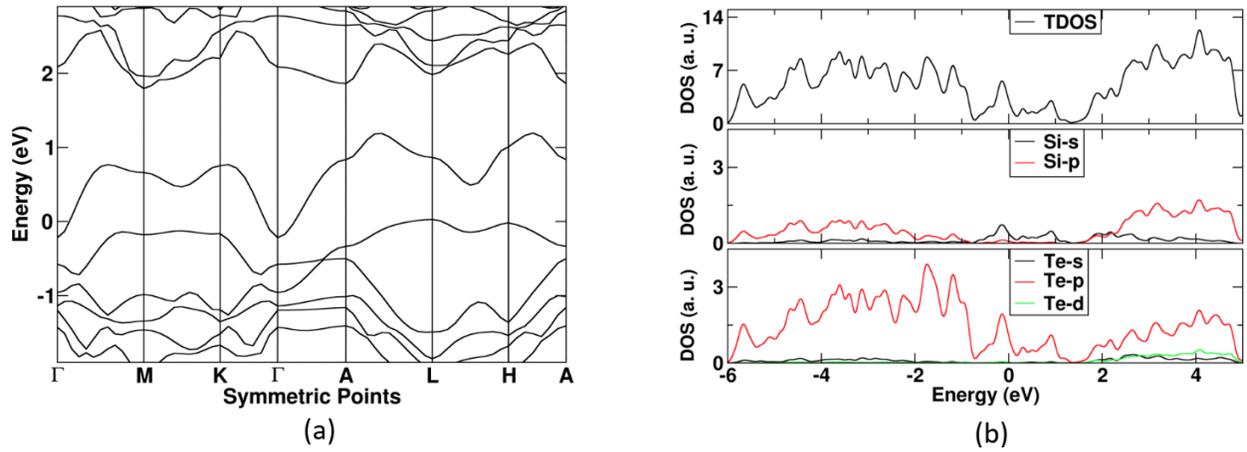

Figure S1. (a) Electronic band structures and (b) DOS of the M1 phase of bulk $Si_2Te_3$ under HSE method.

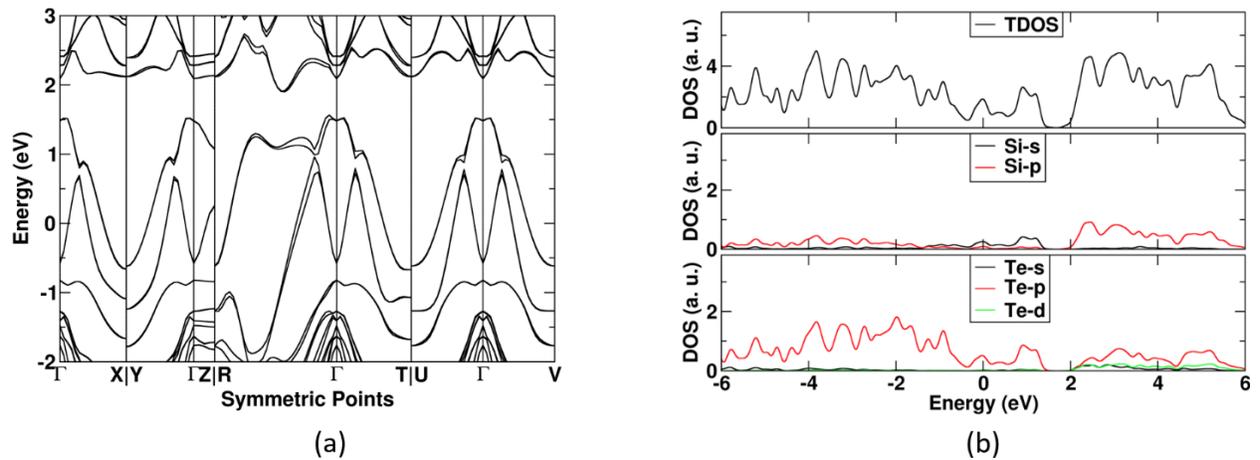

Figure S2. (a) Electronic band structures and (b) DOS of the M2 phase of bulk $Si_2Te_3$ under HSE method.

Figure S1 and S2 represent the electronic band structures and corresponding densities of states (DOS) of the M1 and M2 phases of bulk $Si_2Te_3$ under the hybrid DFT (HSE06) method, respectively. Spin-orbit interaction is considered in the calculations. In both Figures S1 and S2, no band gaps are observed, which is also confirmed by the respective densities of states. These results are consistent with the DFT results, which confirm the metallic behavior of the M1 and M2 phase of bulk $Si_2Te_3$.

The charge density plot of the M1 phase of $Si_2Te_3$ at the Fermi level is shown in Figures S3. It is seen that the Si atoms bonded covalently with the Te atoms (Si1 in Figure 6(a) of the main text) are not contributing to the metallic characteristics. On the other hand, the 2s orbital of the unbonded Si atoms (Si2 in Figure 6(a) of the main text) and the 5p orbital of Te atoms are responsible for states at the Fermi level. These unbonded Si atoms have more charge since they transfer less charge to the Te atoms than the bonded Si atoms, which could be why the unbonded Si atoms are contributing to the metallicity.

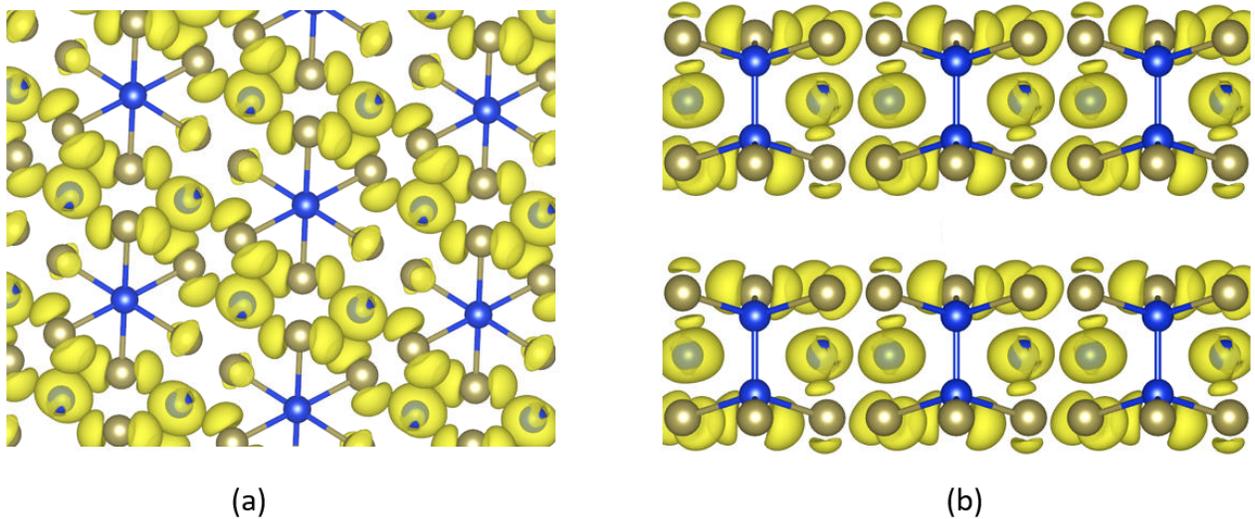

(a)  (b)

Figure S3. Charge density plot of the M1 phase of $Si_2Te_3$ at the Fermi level (a) top view (b) side view. Iso-surface is $1.856 \times 10^{-5}$ in both cases.

Figure S4 shows the charge density plot of the M2 phase of $Si_2Te_3$ at the Fermi level. The 2p-orbital of Si atoms are playing a major role in the conduction at the Fermi level, unlike the 2s orbital in the M1 phase. Also, the Te atoms in the middle layer, which get less charge as compared to the surface atoms, are contributing more than the surface Te atoms.

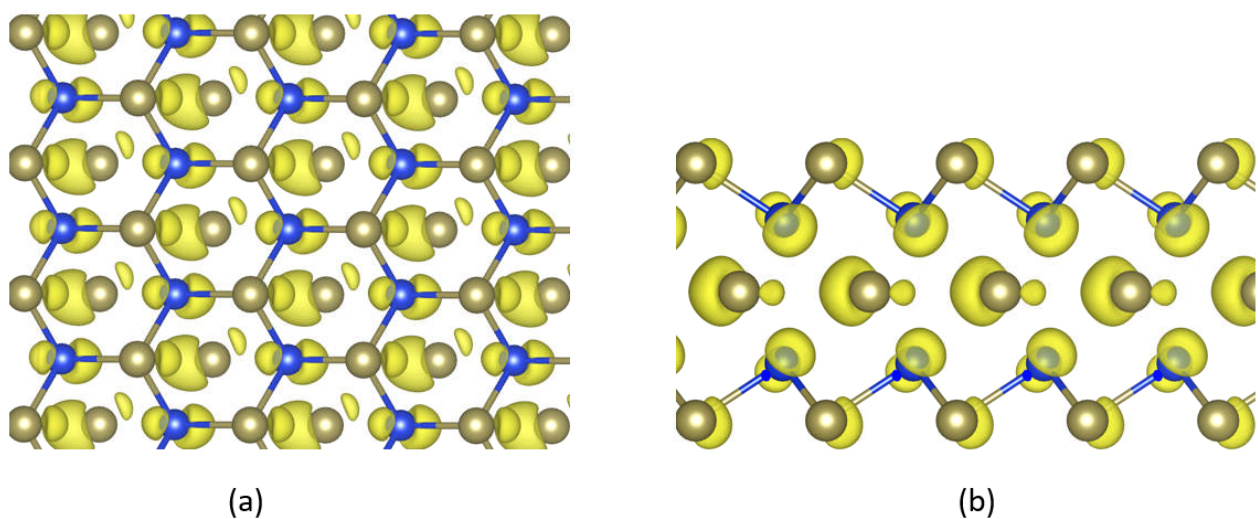

Figure S4. Charge density plot of the M2 phase of Si$_2$Te$_3$ at the Fermi level (a) top view (b) side view. Iso-surface is 1.634 x 10$^{-5}$ in both cases.

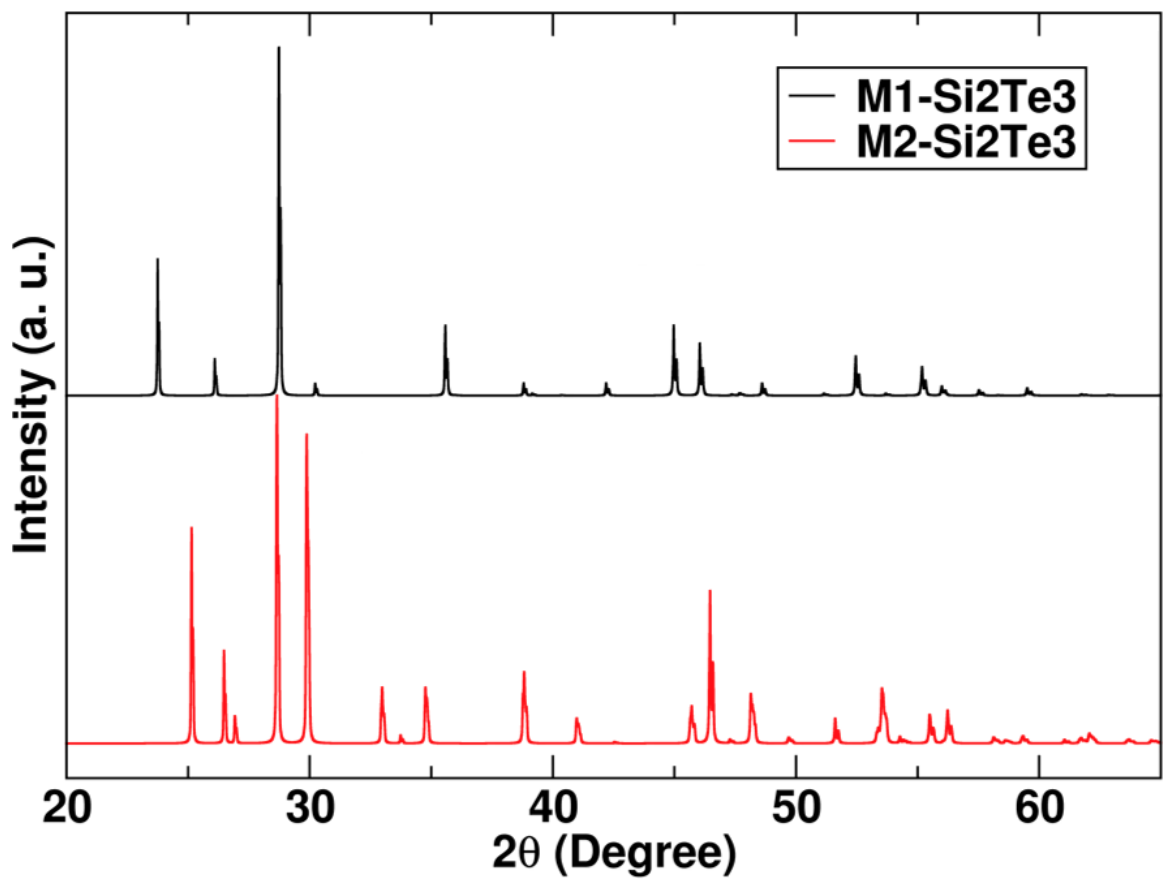

Figure S5. Calculated XRD spectra of the M1 and M2 phases of Si$_2$Te$_3$ using Cu k-alpha radiation showing the major as well as minor Bragg peaks.

Figure S5 shows the calculated X-ray diffraction (XRD) pattern of the M1 and M2 phases of $Si_2Te_3$ under Cu k-alpha method. One major Bragg peak in M1 phase is found at 28.74º, in addition to the minor peaks at 23.75º, 26.11º, 35.58º, 44.97º, 46.05º, 52.45º, and 55.18º. In the M2 phase, two closely spaced major peaks are found at 28.66º and 29.88º, whereas the minor peaks are found at 25.14º, 26.48º, 26.93º, 32.98º, 34.76º, 38.82º, 40.98º, 45.71º, 46.46º, 48.15º, 51.61º, 53.53º, 55.50º, and 56.24º. These major peaks in M1 and M2 phases are close to the experimentally observed peak at 26.5º in SC phase[1]. The prediction of these peaks can be helpful for future experimental investigations of the metallic $Si_2Te_3$.